\documentclass [12pt] {article}
\usepackage{a4}

  \def\pp{{\mathchoice
          {
              \kern 1pt%
              \raise 1pt
              \vbox{\hrule width5pt height0.4pt depth0pt
                    \kern -2pt
                    \hbox{\kern 2.3pt
                          \vrule width0.4pt height6pt depth0pt
                          }
                    \kern -2pt
                    \hrule width5pt height0.4pt depth0pt}%
                    \kern 1pt
           }
            {
              \kern 1pt%
              \raise 1pt
              \vbox{\hrule width4.3pt height0.4pt depth0pt
                    \kern -1.8pt
                    \hbox{\kern 1.95pt
                          \vrule width0.4pt height5.4pt depth0pt
                          }
                    \kern -1.8pt
                    \hrule width4.3pt height0.4pt depth0pt}%
                    \kern 1pt
            }
            {
              \kern 0.5pt%
              \raise 1pt
              \vbox{\hrule width4.0pt height0.3pt depth0pt
                    \kern -1.9pt  
                    \hbox{\kern 1.85pt
                          \vrule width0.3pt height5.7pt depth0pt
                          }
                    \kern -1.9pt
                    \hrule width4.0pt height0.3pt depth0pt}%
                    \kern 0.5pt
            }
            {
              \kern 0.5pt%
              \raise 1pt
              \vbox{\hrule width3.6pt height0.3pt depth0pt
                    \kern -1.5pt
                    \hbox{\kern 1.65pt
                          \vrule width0.3pt height4.5pt depth0pt
                          }
                    \kern -1.5pt
                    \hrule width3.6pt height0.3pt depth0pt}%
                    \kern 0.5pt
            }
        }}

  \def\mm{{\mathchoice
                       {
                             \kern 1pt
               \raise 1pt    \vbox{\hrule width5pt height0.4pt depth0pt
                                  \kern 2pt
                                  \hrule width5pt height0.4pt depth0pt}
                             \kern 1pt}
                       {
                            \kern 1pt
               \raise 1pt \vbox{\hrule width4.3pt height0.4pt depth0pt
                                  \kern 1.8pt
                                  \hrule width4.3pt height0.4pt depth0pt}
                             \kern 1pt}
                       {
                            \kern 0.5pt
               \raise 1pt
                            \vbox{\hrule width4.0pt height0.3pt depth0pt
                                  \kern 1.9pt
                                  \hrule width4.0pt height0.3pt depth0pt}
                            \kern 1pt}
                       {
                           \kern 0.5pt
             \raise 1pt  \vbox{\hrule width3.6pt height0.3pt depth0pt
                                  \kern 1.5pt
                                  \hrule width3.6pt height0.3pt depth0pt}
                           \kern 0.5pt}
                       }}

\catcode`@=11
\def\un#1{\relax\ifmmode\@@underline#1\else
        $\@@underline{\hbox{#1}}$\relax\fi}
\catcode`@=12

\let\du=\du                     

\def\a{\alpha}
\def\b{\beta}

\def\h{\eta}

\def\j{\psi}

\def\m{\mu}

\def\p{\pi}
\def\q{\theta}
\def\r{\rho}

\def\x{\xi}

\def\D{\Delta}

\def\G{\Gamma}

\def\L{\Lambda}
\def\O{\Omega}

\def\Q{\Theta}

\def\ch{{\cal H}}

\def\bo{{\raise-.5ex\hbox{\large$\Box$}}}               
\def\pa{\partial}                                       
\def\de{\nabla}                                         
\def\TH{{\raise.2ex\hbox{$\displaystyle \bigodot$}\mskip-4.7mu \llap H \;}}
\def\face{{\raise.2ex\hbox{$\displaystyle \bigodot$}\mskip-2.2mu \llap {$\ddot
        \smile$}}}                                      

\def\abs#1{\left| #1\right|}                    
\def\leftrightarrowfill{$\mathsurround=0pt \mathord\leftarrow \mkern-6mu
        \cleaders\hbox{$\mkern-2mu \mathord- \mkern-2mu$}\hfill
        \mkern-6mu \mathord\rightarrow$}
\def\dvec#1{\vbox{\ialign{##\crcr
        \leftrightarrowfill\crcr\noalign{\kern-1pt\nointerlineskip}
        $\hfil\displaystyle{#1}\hfil$\crcr}}}           

\def\frac#1#2{{\textstyle{#1\over\vphantom2\smash{\raise.20ex
        \hbox{$\scriptstyle{#2}$}}}}}                   
\def\sfrac#1#2{{\vphantom1\smash{\lower.5ex\hbox{\small$#1$}}\over
        \vphantom1\smash{\raise.4ex\hbox{\small$#2$}}}} 
\def\bfrac#1#2{{\vphantom1\smash{\lower.5ex\hbox{$#1$}}\over
        \vphantom1\smash{\raise.3ex\hbox{$#2$}}}}       
\def\afrac#1#2{{\vphantom1\smash{\lower.5ex\hbox{$#1$}}\over#2}}    

\def\[{\lfloor{\hskip 0.35pt}\!\!\!\lceil}
\def\]{\rfloor{\hskip 0.35pt}\!\!\!\rceil}

\def\du#1#2{_{#1}{}^{#2}}

\def\fracm#1#2{\hbox{\large{${\frac{{#1}}{{#2}}}$}}}

\def\un{\underline}
\def\fracmm#1#2{{{#1}\over{#2}}}

\def\low#1{{\raise -3pt\hbox{${\hskip 0.75pt}\!_{#1}$}}}

\newskip\humongous \humongous=0pt plus 1000pt minus 1000pt
\def\caja{\mathsurround=0pt}
\def\eqalign#1{\,\vcenter{\openup2\jot \caja
        \ialign{\strut \hfil$\displaystyle{##}$&$
        \displaystyle{{}##}$\hfil\crcr#1\crcr}}\,}
\newif\ifdtup

\def\pl#1#2#3{Phys.~Lett.~{\bf {#1}B} (19{#2}) #3}
\def\np#1#2#3{Nucl.~Phys.~{\bf B{#1}} (19{#2}) #3}

\parskip=\medskipamount                 
\thicklines                         

\begin{document}
\thispagestyle{empty}

{\hbox to\hsize{
\vbox{\noindent February 2003 \hfill hep-th/0302001 }}}

\noindent
\vskip1.3cm
\begin{center}

{\Large\bf D-instantons and matter hypermultiplet}

\vglue.2in

                  Sergei V. Ketov

{\it          Department of Physics\\
           Tokyo Metropolitan University\\
          Hachioji, Tokyo 192--0397, Japan}
\vglue.1in
{\sl ketov@phys.metro-u.ac.jp}

\end{center}

\vglue.3in

\begin{center}
{\Large\bf Abstract}
\end{center}

\noindent
We calculate the D-instanton corrections (with all D-instanton numbers) to the
quantum moduli space metric of a single matter hypermultiplet with toric 
isometry, in the effective N=2 supergravity arising in type-IIA superstrings 
compactified on a Calabi-Yau (CY) threefold of Hodge number $h_{2,1}=1$. The 
non-perturbative quaternionic hypermultiplet metric is derived by resolution 
of a complex orbifold singularity, thus generalizing the known (Ooguri-Vafa) 
solution in flat spacetime to N=2 supergravity.

\newpage

\section{Introduction}

D-instanton quantum corrections to the moduli space metric of a single
matter hypermultiplet for the CY-compactified type IIA superstrings near a 
conifold singularity were investigated by Ooguri and Vafa \cite{ov}. They 
found an unique solution consistent with N=2 {\it rigid} supersymmetry and 
toric isometry. The solution \cite{ov} was interpreted as the infinite 
D-instanton sum coming from multiple wrappings of the Euclidean D-branes 
around the vanishing cycle \cite{ov}. The Ooguri-Vafa solution is given by the 
hyper-K\"ahler metric in the limit of flat 4d spacetime, i.e. when N=2 
supergravity decouples, by taking the Planck mass to infinity. Including N=2 
supergravity may lead to new physical phenomena at strong coupling 
\cite{bbs}, so that it is important to generalize the Ooguri-Vafa solution to 
the case of {\it local} N=2 supersymmetry.

One immediate consequence of the local N=2 supersymmetry is that the 
hypermultiplet moduli space metric is no longer hyper-K\"ahler but 
quaternionic \cite{bw}. In particular, the standard (Gibbons-Hawking)  
 hyper-K\"ahler Ansatz  \cite{gh} (it was one of the key technical tools 
in ref.~\cite{ov}) cannot be applied to quaternionic metrics. Hence, the 
quaternionic generalization of the Ooguri-Vafa solution is a non-trivial
problem.

In our earlier papers \cite{my} we solved a similar problem in the case of the
{\it universal} hypermultiplet that has a dilaton amongst its bosonic field
components. The D-instanton corrections to the universal hypermultiplet moduli
space metric were derived in ref.~\cite{my} by requiring, in addition to N=2
local supersymmetry and toric isometry, the $SL(2,{\bf Z)}$ duality invariance
that is not the case for a single matter hypermultiplet. The moduli space 
metric of the matter hypermultiplet is periodic (or T-selfdual) with respect 
to one of its variables \cite{ov} but it is not S-selfdual, so that another 
solution is needed. 

The quaternionic constraints in the  case of a single hypermultiplet are 
given by the (integrable) Einstein-Weyl system of non-linear partial 
differential equations. The key to the explicit construction of ref.~\cite{my}
 is the remarkable fact that any Einstein-Weyl metric with toric isometry is 
governed by the real pre-potential that is an eigenfunction of the 
Laplace-Beltrami operator $\D_{\ch}$ in the hyperbolic plane $\ch$ with the
eigenvalue $3/4$ \cite{cp}. The four-dimensional hyper-K\"ahler metrics 
(including the Ooguri-Vafa metric \cite{ov}) with a triholomorphic isometry 
are known to be governed by {\it harmonic} functions in flat three dimensions
(except some isolated points) \cite{gh}. In this Letter we also use yet 
another mathematical fact that the eigenfunctions of $\D_{\ch}$ correspond to 
the homogeneous harmonic functions \cite{cs}. The solution \cite{ov} is 
formally obtained by applying T-duality to the `basic' Green function in three
dimensions.  By using the Calderbank-Singer correspondence \cite{cs} we can 
lift the Ooguri-Vafa solution \cite{ov} to a curved spacetime of N=2 
supergravity by applying T-duality to the corresponding basic eigenfunction of
 $\D_{\ch}$. In the next sect.~2 we review the hyper-K\"ahler solution 
\cite{ov} that is the pre-requisite to our derivation of the corresponding 
quaternionic solution in sect.~3. 
 
\section{Ooguri-Vafa solution}

The {\it Ooguri-Vafa} (OV) solution \cite{ov} describes the D-instanton 
corrected moduli space metric of a single matter hypermultiplet in N=2 
four-dimensional superstrings compactified on a Calabi-Yau threefold of Hodge 
number $h_{2,1}=1$, when both N=2 supergravity and the universal 
hypermultiplet are switched off, and five-brane instantons are suppressed. 
The corresponding low-energy effective action is given by the four-dimensional
 N=2 supersymmetric non-linear sigma-model that has the Ooguri-Vafa metric in 
its target space.

Rigid N=2 supersymmetry of the non-linear sigma-model requires a 
hyper-K\"ahler metric \cite{fal,book}. The OV metric also has a 
(toric) $U(1)\times U(1)$ isometry by construction \cite{ov}. There always 
exist a linear combination of two commuting abelian isometries that is 
tri-holomorphic, i.e. it commutes with N=2 rigid supersymmetry \cite{gib}.

Given any four-dimensional hyper-K\"ahler metric with a tri-holomorphic 
isometry $\pa_t$, it can always be written down in the standard 
(Gibbons-Hawking) form \cite{gh},
$$ ds^2_{\rm GH}= \fracmm{1}{V}(dt+\hat{\Q})^2 +V(dx^2+dy^2+dz^2)~, 
\eqno(1)$$
that is governed by {\it linear} equations,
$$ \D V=\vec{\nabla}{}^2V\equiv \left( \fracmm{\pa^2}{\pa x^2}+
\fracmm{\pa^2}{\pa y^2}
+\fracmm{\pa^2}{\pa z^2}\right)V=0~,\quad {\rm almost~~everywhere},\eqno(2)$$
and
$$\vec{\de}V+\vec{\de}\times\vec{\Q}=0~.\eqno(3)$$
The one-form $\hat{\Q}=\Q_1dx +\Q_2dy+\Q_3dz$ is fixed by the 
`monopole equation' (3) in terms of the real scalar potential $V(x,y,z)$. 
Equation (2) means that the function $V$ is harmonic, with possible
 isolated singularities, in three Euclidean dimensions ${\bf R}^3$. The 
singularities are associated with the positions of D-instantons.

Given extra $U(1)$ isometry, after being rewritten in the cylindrical 
coordinates ($\r=\sqrt{x^2+y^2},~\theta=\arctan(y/x),~\h=z)$, the 
hyper-K\"ahler potential $V(\r,\q,\h)$ becomes independent upon $\theta$. 
Equation (1) was used by Ooguri and Vafa \cite{ov} in their analysis of the 
matter hypermultiplet moduli space near a conifold singularity. The conifold 
singularity arises in the limit of the vanishing period, 
$$ \int_{\cal C}\O\to 0~~,\eqno(4)$$
where the CY holomorphic 3-form $\O$ is integrated over a non-trivial 3-cycle 
${\cal C}$ of CY. The powerful singularity theory \cite{sin}  can then be 
applied to study the universal behaviour of the hypermultiplet moduli space  
near the conifold limit. 

In the context of the CY compactification of type IIA superstrings, the
coordinate $\r$ represents the `size' of the CY cycle ${\cal C}$ or, 
equivalently, the action of the D-instanton originating from the Euclidean 
D2-brane wrapped about the cycle ${\cal C}$. The physical interpretation of 
the $\h$ coordinate is just the expectation value of another (RR-type) 
hypermultiplet scalar. The cycle ${\cal C}$ can be replaced by a sphere $S^3$
for our purposes, since the D2-branes only probe the overall size of 
${\cal C}$, as in ref.~\cite{ov}.

The pre-potential $V$ is {\it periodic} in the RR-coordinate $\h$ since 
the D-brane charges are quantized \cite{bbs}. This periodicity should also be 
valid in curved spacetime. We normalize the corresponding period to be $1$, 
as in ref.~\cite{ov}. The Euclidean D2-branes wrapped $m$ times around the 
sphere $S^3$ couple to the RR  expectation value on $S^3$ and thus should 
produce the additive contributions to the pre-potential $V$, with the factor 
of $\exp(2\p im\h)$ each.

In the {\it classical} hyper-K\"ahler limit, when both N=2 supergravity and 
all D-instanton contributions are suppressed, the pre-potential $V(\r,\h)$ of
a single matter hypermultiplet cannot depend upon $\h$ since there is no 
perturbative superstring state with a non-vanishing RR charge. Accordingly, 
the classical pre-potential $V(\r)$ can only be the Green function of the 
two-dimensional Laplace operator, i.e. 
$$ V_{\rm classical} = -\fracmm{1}{2\p}\log\r + {\rm const.}~,\eqno(5)$$
whose normalization is in agreement with ref.~\cite{ov}. 

The calculation of ref.~\cite{ov} to determine the exact D-instanton 
contributions to the hyper-K\"ahler potential $V$ is based on the idea 
\cite{bbs} that the D-instantons should resolve the singularity of the 
classical hypermultiplet moduli space metric at $\r=0$. A similar situation 
arises in the standard (Seiberg-Witten) theory of a quantized N=2 vector 
multiplet (see, e.g., ref.~\cite{book} for a review).

Equation (2) formally defines the electrostatic potential $V$ of electric 
charges of unit charge in the Euclidean upper half-plane $(\r,\h)$, $\r>0$, 
which are distributed along the axis $\r=0$ in each point $\h=n\in {\bf Z}$, 
while there are no two charges at the same point \cite{ov}. A solution to 
eq.~(2) obeying all these conditions is unique, 
$$ V\low{\rm OV}(\r,\h)= \fracmm{1}{4\p} \sum^{+\infty}_{n=-\infty}\left(
\fracmm{1}{\sqrt{\r^2+ (\h-n)^2}}-\fracmm{1}{\abs{n}}\right)+{\rm const.}
\eqno(6)$$
After the Poisson resummation eq.~(6) takes the singularity resolution form 
indeed \cite{ov},
$$ V\low{\rm OV}(\r,\h)=\fracmm{1}{4\p} \log\left( \fracmm{\m^2}{\r^2}\right)+
\sum_{m\neq 0}\fracmm{1}{2\p}e^{2\p im\h}\,K_0\left(2\p \abs{m}\r\right)~,
\eqno(7)$$
where the modified Bessel function $K_0$ of the 3rd kind has been introduced,
$$ K_s(z)=\fracm{1}{2}\int^{+\infty}_0\fracmm{dt}{t^{s-1}}\exp\left[
-\,\fracmm{z}{2}\left( t+\fracmm{1}{t}\right)\right]~,\eqno(8)$$
valid for all Re\,$z>0$ and  Re\,$s>0$, while $\m$ is a constant.

By inserting the standard asymptotical expansion of the Bessel function $K_0$ 
near $\r=\infty$ into eq.~(7) one finds \cite{ov}
$$\eqalign{
 V\low{\rm OV}(\r,\h)~=~&\fracmm{1}{4\p} \log 
\left( \fracmm{\m^2}{\r^2}\right) +
\sum_{m=1}^{\infty} \exp \left(-2\p m\r\right)
\cos(2\p m\h)\times\cr
~& \times \sum_{n=0}^{\infty}\fracmm{\G(n+\fracm{1}{2})}{\sqrt{\p}n!
\G(-n+\fracm{1}{2})}\left(\fracmm{1}{4\p m\r}
\right)^{n+\frac{1}{2}}~~~.\cr}\eqno(9)$$

A dependence upon the string coupling constant $g_{\rm string}$ is easily
reintroduced into eq.~(9) by a substitution $\r\to\r/g_{\rm string}$.
The factors of $\exp{(-2\p m\r/g_{\rm string})}$ in eq.~(9) are just the 
expected multi-D-instanton contributions \cite{ov}. 

The OV pre-potential (6) has the form of the (regularized) T-sum over the
 T-duality transformations, $\h\to \h+1$, applied to the `basic' solution 
$V_0\equiv\fracmm{1}{4\p r}\equiv \fracmm{1}{4\p\sqrt{\r^2+\h^2}}$ of eq.~(2),
$$ V\low{\rm OV}(\r,\h)=A + \sum_{\rm T}V_0(\r,\h)= A + \sum_{\rm T} 
\fracmm{1}{4\p\sqrt{\r^2+\h^2}}~~~,\eqno(10)$$
where $A$ is a constant. The basic solution $V_0(\r,\h)$ is just the Green 
function of the three-dimensional Laplace operator $\D$ in eq.~(2).

\section{D-instantons in N=2 supergravity}

Any four-dimensional quaternionic manifold has the Einstein-Weyl geometry of 
 negative scalar curvature \cite{bw},
$$ W^-_{abcd}=0~,\qquad R_{ab}=-\fracmm{\L}{2}g_{ab}~,\qquad
a,b,c,d=1,2,3,4~,\eqno(11)$$  
where $W_{abcd}$ is the Weyl tensor, $R_{ab}$ is the Ricci tensor of the 
metric $g_{ab}$, and the constant $\L>0$ is proportional to the gravitational 
coupling constant.

It is the theorem \cite{cp} that {\it any} four-dimensional Einstein-Weyl
metric (of non-vanishing scalar curvature) with two linearly independent 
Killing vectors can be written down in the form 
$$\eqalign{ 
ds^2_{\rm CP} ~=~ &  \fracmm{4\r^2(F^2_{\r}+F^2_{\h})-F^2}{4F^2}\,
\left(\fracmm{d\r^2+d\h^2}{\r^2}\right) \cr 
 & + \fracmm{ [(F-2\r F_{\r})\hat{\a}-2\r F_{\h}\hat{\b} ]^2 +[-2\r F_{\h}
\hat{\a}
+(F+2\r F_{\r})\hat{\b}]^2 }{F^2[4\r^2(F^2_{\r}+F^2_{\h})-F^2] }~,\cr}
\eqno(12)$$
in some local coordinates $(\r,\h,\q,\j)$ inside an open region of the 
half-space $\r>0$, where $\pa_{\q}$ and $\pa_{\j}$ are the two Killing 
vectors, the one-forms $\hat{\a}$ and $\hat{\b}$ are given by
$$ \hat{\a}= \sqrt{\r}\,d\q\quad {\rm and}\quad \hat{\b}=\fracmm{d\j 
+\h d\q}{\sqrt{\r}}~~,\eqno(13)$$
while the whole metric (12) is governed by the real function 
(= {\it hypermultiplet pre-potential}) $F(\r,\h)$ obeying a linear 
differential equation,
$$\D_{\ch}F \equiv \r^2\left(\pa^2_{\r}+\pa^2_{\h}\right)F =
\fracmm{3}{4}F~~.\eqno(14)$$

Equation (14) is thus a consequence of 4d, local N=2 supersymmetry and 
toric isometry. It is highly non-trivial that the {\it linear} master equation 
(14) governs all $U(1)\times U(1)$-symmetric solutions to the highly 
{\it non-linear} Einstein-Weyl system (11).

Equation (14) means that the quaternionic pre-potential $F$ is a 
local eigenfunction (of the eigenvalue $3/4$) of the two-dimensional 
$SL(2,{\bf R})$ Laplace-Beltrami operator 
$$\D_{\ch} =\r^2(\pa^2_{\r}+\pa^2_{\h}) \eqno(15)$$
on the hyperbolic plane $\ch$ equipped with the metric 
$$ ds^2_{\ch}= \fracmm{1}{\r^2}( d\r^2 +d\h^2)~,\quad \r > 0~. 
\eqno(16)$$  

The `basic' $\h$-independent solutions to eq.~(14) are given by
$$ \r^{-1/2}\qquad {\rm and}\qquad \r^{3/2}~.\eqno(17a)$$
Their linear combination gives some perturbative contributions to the 
hypermultiplet pre-potential \cite{my}. The only $\h$-dependent `basic' 
solution to eq.~(14) is given by \cite{cp,cs}
$$ F_0(\r,\h)=\sqrt{ \r +\fracmm{\h^2}{\r}}~~~.\eqno(17b)$$
A linear combination of those solutions with T-shifts $\h\to \h+n$ is known to
describe a generic multi-instanton quaternionic metric via eqs.~(12) and 
(14) \cite{cp,my}.

The basic eigenfunction (17b) is simply related to the harmonic (outside the 
origin) function $V_0$ (see the end of sect.~2),
$$4\p V_0(\r,\h)= \fracmm{\pa(\sqrt{\r}F_0)}{\r\pa\r}~~.\eqno(18)$$
This is an example of the general correspondence between the harmonic functions
$V$ of homogeneity $\a$ in ${\bf R}^3$ and the $\D_{\ch}$-eigenfunctions $F$ 
in $\ch$ of eigenvalue $\a(\a-1)$ \cite{cs}.

Applying T-duality (i.e. summing up over the orbit with respect to the 
T-transformation $\h\to \h+1$) to the linear relation (18) yields
$$ V_{\rm OV}(\r,\h)=A+\sum_{\rm T}V_0(\r,\h) =
\fracmm{\pa(\sqrt{\r}\sum_{\rm T}F_0)}{4\p\r\pa\r}
 =\fracmm{\pa(\sqrt{\r}F)}{4\p\r\pa\r}~~.\eqno(19)$$
The quaternionic hypermultiplet potential $F$, corresponding to the Ooguri-Vafa
 solution $ V_{\rm OV}$, is thus given by eq.~(19). A general solution to 
eq.~(19) reads 
$$ \sqrt{\r}F(\r,\h)=4\p \int^{\r}_0d\x~\x\left[ V_{\rm OV}(\x,\h)+B \right]
 +f(\h) \eqno(20)$$
with some function $f(\h)$, and a constant $B$. Substituting eq.~(7) 
into eq.~(20) and using the identities
$$ \fracmm{d}{dx}\left[xK_1(x)\right] = -xK_0(x)~,\quad
\lim_{x\to 0} xK_1(x)=1~,\quad
\sum^{\infty}_{k=1}\fracmm{\cos(kx)}{k^2}=\fracmm{\p^2}{6}-\fracmm{\p x}{2}+
\fracmm{x^2}{4}~,\eqno(21)$$
we find
$$ \sqrt{\r}F(\r,\h)=\tilde{f}(\h)+
\fracmm{\r^2}{2}\ln\fracmm{\tilde{\m}^2}{\r^2} -
\fracmm{2\r}{\p}\sum_{k=1}^{\infty}
\fracmm{\cos(2\p k\h)}{k}K_1(2\p k\r)~,\eqno(22)$$
where we have redefined the integration function $f(\h)$ and the 
renormalization parameter $\m$ as
$$ \tilde{f}(\h)=f(\h) + \fracmm{1}{6}\left[ 1+ 6\h(\h-1)\right]~,\qquad
1+4\p B +\ln\m^2=\ln\tilde{\m}^2~.\eqno(23)$$

In the perturbative region at large $\r$ the sum in eq.~(22) represents the 
non-perturbative contributions with all D-instanton numbers, when  using the 
asymptotical expansion of the $K_1$-function at $x\to\infty$, 
$$ K_1(x)\sim \sqrt{\fracmm{\p}{2x}}\,e^{-x}\left[1+{\cal O}(x^{-1})\right]~,
\eqno(24)$$
in full similarity to ref.~\cite{ov}. The first two terms in eq.~(22) are
perturbative contributions that are to be $\h$-independent. Hence, the
function $\tilde{f}(\h)$ is actually a constant, $\tilde{f}=C$. Our final 
result for the D-instanton-corrected pre-potential of the matter hypermultiplet
 moduli space metric is thus given by
$$ F(\r,\h)=\fracmm{C}{\sqrt{\r}}+ \fracmm{\r^{3/2}}{2}
\ln\fracmm{\tilde{\m}^2}{\r^2} -\fracmm{2\sqrt{\r}}{\p}\sum_{k=1}^{\infty}
\fracmm{\cos(2\p k\h)}{k}K_1(2\p k\r)~.\eqno(25)$$
The logarithmic factor in the second term of eq.~(25) is apparently to be 
interpreted as the N=2 superstring (one-loop) renormalization effect.

I am grateful to the referee of my paper \cite{my} in Nucl. Phys. B, whose 
remarks led me to this investigation.


\begin{thebibliography}{99}
\bibitem{ov} H. Ooguri and C. Vafa, Phys. Rev. Lett. {\bf 77} (1996) 3298 
[hep-th/9608079] 
\bibitem{bbs} K. Becker, M. Becker and A. Strominger, \np{456}{95}{130} 
\newline [hep-th/9507158] \\
A. Strominger, \pl{421}{98}{139} [hep-th/9706195] \\
K. Becker and M. Becker, \np{551}{99}{102} [hep-th/9901126]
\bibitem{bw} J. Bagger and E. Witten, \np{222}{83}{1}
\bibitem{gh} G. W. Gibbons and S. W. Hawking, Phys. Lett. {\bf 78B} (1978) 430 
\bibitem{my} S. V. Ketov, {\it D-Instantons and universal hypermultiplet},
CITUSC preprint 01--046 [hep-th/0112012], unpublished; 
Fortschr. Phys. {\bf 50} (2002) 909 [hep-th/0111080]; Nucl. Phys. {\bf B649} 
(2003) 365 [hep-th/0209003]
\bibitem{cp} D. M. J. Calderbank and H. Pedersen, J. Diff. Geom. {\bf 60}
 (2002) 485 [math.DG/0105263]
\bibitem{cs} D. Calderbank and M. A. Singer, {\it Einstein metrics and complex
singularities}, math.DG/0206299
\bibitem{fal} L. Alvarez-Gaum\'e and D. Z. Freedman, Commun. Math. Phys. 
{\bf 80} (1981) 443 
\bibitem{book} S. V. Ketov, {\it Quantum Non-linear Sigma-Models}, 
Springer-Verlag, 2000 
\bibitem{gib} G.W. Gibbons, D. Olivier, P. J. Ruback, G. Valent,  
Nucl. Phys. {\bf B296} (1988) 679
\bibitem{sin} V. Arnold, A. Gusein-Zade and A. Varchenko, {\it Singularities
of Differentiable Maps}, Birkh\"auser, 1985.

\end{thebibliography}
\end{document}